\documentclass[12pt]{article}
\usepackage{graphicx}
\begin{document}
\centerline{\Large \bf On the Consensus Threshold for the}
\centerline{\Large \bf Opinion Dynamics of Krause-Hegselmann}

\bigskip
\bigskip

\centerline{Santo Fortunato}

\vskip0.3cm

\centerline{Fakult\"at f\"ur Physik, Universit\"at Bielefeld, 
D-33501 Bielefeld, Germany}

\noindent

\centerline{{\tt e-mail: fortunat@physik.uni-bielefeld.de}}

\bigskip

\begin{abstract}

In the consensus model of Krause-Hegselmann, 
opinions are real numbers between $0$ and $1$ and two agents
are compatible if the difference of their opinions is smaller 
than the confidence bound parameter $\epsilon$.
A randomly chosen agent takes the average of the opinions of all neighbouring 
agents which are compatible with it. 
We propose a conjecture, based on numerical evidence, on the value
of the consensus threshold $\epsilon_c$ of this model. 
We claim that $\epsilon_c$ can take only two possible
values, depending on the 
behaviour of the average degree $d$ of the graph representing
the social relationships, when the population $N$ goes to infinity:
if $d$ diverges when $N\rightarrow\infty$, $\epsilon_c$ equals the consensus
threshold $\epsilon_i\sim\,0.2$ on the complete graph; if instead
$d$ stays finite when $N\rightarrow\infty$, $\epsilon_c=1/2$ as for the 
model of Deffuant et al.

\end{abstract}

\bigskip

Keywords: Sociophysics, Monte Carlo simulations.

\bigskip

In recent years several models of opinion formation have been proposed \cite{Axel,
Deff, HK, Sznajd, Galam,huberman}. In general they deal with simple cellular automata, where 
people become the vertices of a graph and neighbouring vertices represent
agents which have a personal relationship (acquaintance). A simple rule 
determines how the opinion of an agent is influenced from (or can influence)
that of its neighbours. The aim is to understand how it happens that large 
groups of people ultimately share the same opinion, starting from a situation
in which everybody has its own ideas independently of those of the people with whom
they interact. In particular, for large-scale phenomena it is possible
to predict and/or reproduce general features like statistical distributions:
with a voter model based on the Sznajd \cite{Sznajd} dynamics one was able to reproduce
the final distribution of votes among 
candidates in Brazilian and Indian elections \cite{st,gonz}.

In this paper we focus on a special model, that of Krause-Hegselmann (KH)
\cite{HK}. It is a compromise model based on the principle of bounded confidence,
which characterizes as well the opinion dynamics of Deffuant et al. \cite{Deff}.
Bounded confidence is nothing but the reasonable consideration 
that a discussion between two individuals is constructive, i.e. it may lead
to a change of opinion, only if the initial positions of the two persons
are close enough, otherwise everybody retains its own opinion. 
In the model of KH, the opinion $s$ is a real number in [0:1],
and two opinions are 'close' to each other if the absolute value of their difference 
is smaller than a positive real parameter $\epsilon$, called confidence bound.

One starts from a graph $G$ and assigns to each of its vertices a real number between 
$0$ and $1$, 
with uniform probability. Next, we perform ordered sweeps through the whole system and
iteratively update the opinion of each agent. 
Suppose we want to update the opinion of an agent $i$:
one has to check which agents among the neighbours of $i$ have opinions which are 'close'
to that of $i$, in the sense explained above. Only such neighbours, that span some set $\cal V$,
can influence the opinion of $i$. The new opinion $s_i$ of $i$
is given by the average\footnote{We remark 
that one could put the agent $i$ itself in the set $\cal V$, so that the opinion
$s_i$ would contribute to the average too. This fact
would be reasonable because the 
initial opinion should somehow be taken into account in the decision 
process of the individual, but it would have no influence whatsoever on the final 
configurations attained by the system.} of the opinions 
of the agents in $\cal V$. By repeating the procedure over and over, the system 
will reach a configuration which is a fixed point for the dynamics, so it is stable.
Such configuration is characterized by just a few surviving opinions, with many agents sharing 
the same opinion. Strictly speaking, a stable configuration must be a superposition of 
Dirac $\delta$'s, located in such a way that the distance between two consecutive
spikes is larger than $\epsilon$.   

The number of opinion clusters in the final configuration depends on the
value of $\epsilon$. In particular, above some value $\epsilon_c$, all agents are bound to 
share the same opinion at the end of the process (consensus). 
The location of this threshold for complete synchronization is very important, because it
provides useful information on the dynamics of the model. For the model of Deffuant et al.,
for instance, we have recently discovered that the threshold for complete consensus is
$\epsilon_c=1/2$, independently of the particular graph used to modelize society \cite{santo}.
In the model of Deffuant et al. \cite{Deff} 
the interactions between the agents are binary processes: an individual
chooses at random one of its neighbours and discusses with it. If the two neighbours are compatible,
i. e. if their opinions differ from each other (in absolute value) by less than 
the confidence bound $\epsilon$, the opinions of the agents move towards each other 
by a relative amount $\mu$, where $\mu$ is real in $[0:1/2]$.

The model of KH was originally introduced 
for a community where everybody talks to everybody else \cite{HK},
and in this special case the algorithm runs very slowly as compared to that of Deffuant et al., 
due to the large averages needed to update the opinions of the agents. In this way, 
at present one can simulate at most systems with populations of the 
order of $10^5$ agents, whereas with the Deffuant dynamics systems as large as the 
population of the European Union can be simulated \cite{staufrev}. That is the
main reason why much less is known \cite{san1,san2} on 
the model of KH as compared to that of Deffuant et al.

Nevertheless, we believe that the model of KH deserves more attention from the sociophysics 
community. In a sense, it is similar to the model of Deffuant et al., as it
follows the criterion of bounded confidence and the
opinion of an agent is affected by that of its neighbours in an "average" way:
the crucial difference is that in KH the agent feels in one shot the influence
of all its (compatible) neighbours, in Deffuant this happens after some time because each
interaction of the agent involves only one of its (compatible) 
acquaintances. This argument suggests
that there may be a considerable difference between the two models when 
the number of neighbours, or {\it degree}, of each agent is large,
but that this difference should reduce when
the degree is small. Moreover, in the latter case it takes just a short
time to calculate the average opinion of the compatible neighbours, because
there are only a few, and the algorithm can compete in speed with that of
Deffuant. Indeed, for special graphs like regular lattices or random graphs with
low average degree, which are much more appropriate for realistic applications,
the KH algorithm is faster than the Deffuant algorithm, except perhaps 
in the very narrow bands of $\epsilon$ corresponding to the transition from a
stable final opinion configuration to the next one, where the dynamics slows down.

Here we will investigate the consensus threshold of the KH model, by performing
a similar analysis as in \cite{santo}. Basically, we carried
out simulations of the model on different graph topologies and determined in each
case the value of the consensus threshold. It turns out that the scenario is 
more complex than for the model of Deffuant et al., as we expected, but that the
consensus threshold keeps its character of universality, labeling large classes
of graphs. We analyzed five different types of graphs:

\begin{itemize}
\item{a complete graph, where everybody talks to everybody else \cite{HK};}
\item{a square lattice;}
\item{a scale free graph {\'a} la Barab{\'a}si-Albert \cite{BA};}
\item{a random graph {\'a} la Erd{\"o}s and R{\'e}nyi \cite{erdos};}
\item{a star-like graph where a vertex is connected to all the others and no
    further connections exist.}
\end{itemize}

\begin{figure}[hbt]
\begin{center}
\includegraphics[angle=-90,scale=0.45]{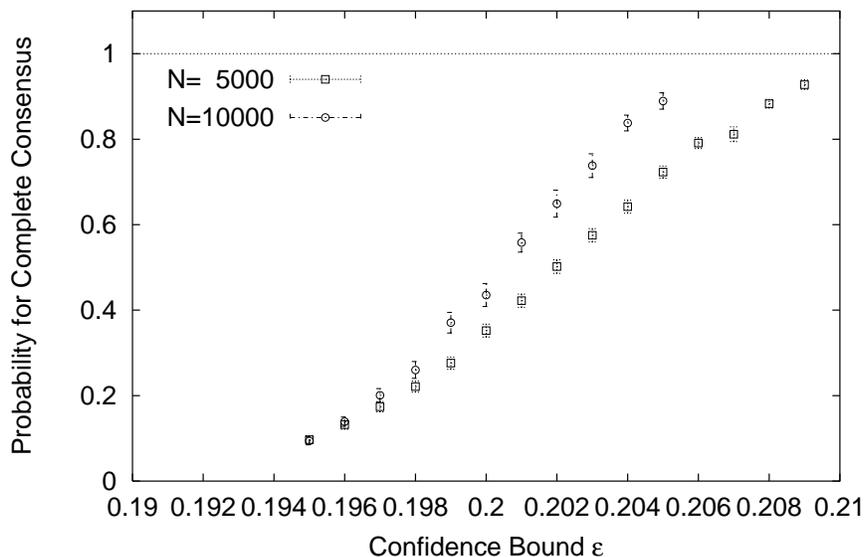}
\end{center}
\caption{\label{fig1}Fraction of samples with a single opinion cluster 
in the final configuration, for a society where everybody talks to
everybody. The two data sets refer to a population of $5000$ and $10000$ agents.}
\end{figure}

Before discussing the single cases, we give some details on the Monte Carlo simulations.
We chose to update the opinions of the agents
in ordered sweeps over the population. The equally legitimate choice of random 
updating would not have influence on the final number of opinion
clusters\footnote{Nevertheless for special types of graphs this choice
  can influence the probability for complete consensus (see Figs. \ref{fig6} and
  \ref{fig7}).}.
The program stops if no agent changed opinion after an iteration; since 
opinions are $64$-bit real numbers, our criterion is to check whether any
opinion varied by less than $10^{-9}$ after a sweep.
We proceeded as follows: for a given population $N$ 
and confidence bound $\epsilon$
we produced $1000$ configurations. 
After that we analyzed the final configurations, by checking 
whether all agents are labeled by the same opinion variable or not ("the same"
means still within $10^{-9}$).
The fraction of samples with all agents sharing the same opinion is the 
probability $P_c$ to have complete consensus, that we study as a function of
$\epsilon$. For each social topology we repeated the procedure 
for several values of the population size,
because the scaling of the curves with $N$ is useful
to better identify the position of the consensus threshold.

\begin{figure}[hbt]
\begin{center}
\includegraphics[angle=-90,scale=0.45]{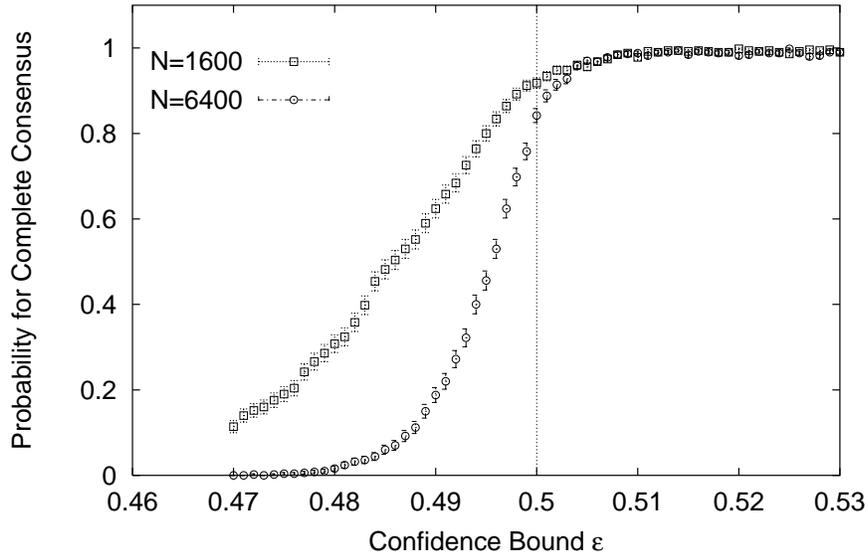}
\end{center}
\caption{\label{fig2}As Fig. \ref{fig1}, but for agents sitting on the sites of
  a square lattice with periodic boundary conditions.
The two data sets refer to a population of $1600$ and $6400$ agents.}
\end{figure}

We present our results starting from the case of the complete graph. 
Fig. \ref{fig1} shows how the consensus probability $P_c$ varies as a function
of $\epsilon$. The two curves correspond to a population of $5000$ 
and $10000$ agents, respectively. For the reasons we explained above, it is
virtually impossible
to go to much larger values of $N$, as the algorithm would become 
terribly slow. Nevertheless, from Fig. \ref{fig1} we observe that $P_c$ rapidly 
rises in a rather narrow interval of the opinion space. From the figure 
it seems that the curve will approach a step function, as already observed in \cite{santo}.
We cannot determine with precision the position of the step in the limit where
the population $N$ goes to infinity, but it will almost surely lie within the 
observed variation range
$[0.195, 0.202]$. We set the upper limit to $0.202$ because for $\epsilon>0.202$
the curve corresponding to $N=10000$ is definitely above the curve relative to
$N=5000$, which suggests that when $N$ diverges $P_c$ will probably attain the
value $1$ in that region. So, we conclude that the consensus threshold
for KH on the complete graph, that we indicate with $\epsilon_i$, is in the
interval $[0.195, 0.202]$.

\begin{figure}[hbt]
\begin{center}
\includegraphics[angle=-90,scale=0.45]{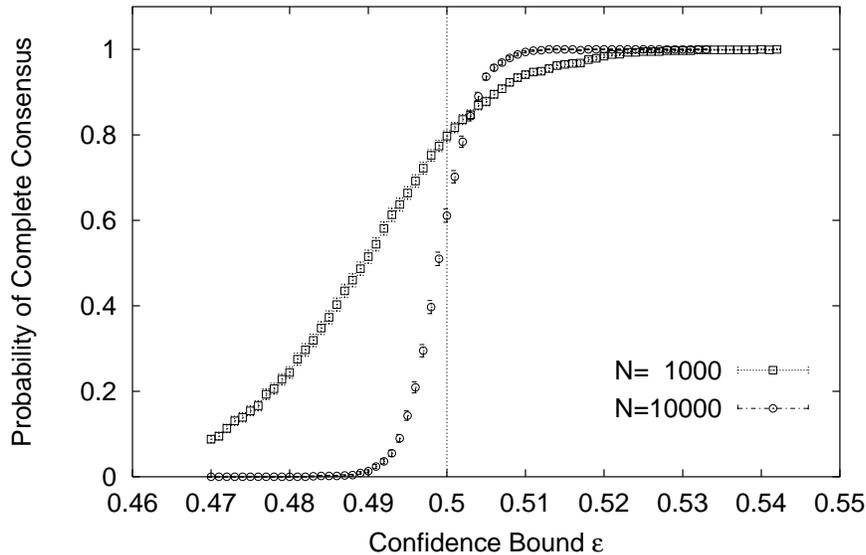}
\end{center}
\caption{\label{fig3}As Fig. \ref{fig1}, but for agents sitting on the sites of
  a scale free network {\'a} la Barab{\'a}si-Albert.
The two data sets refer to a population of $1000$ and $10000$ agents.}
\end{figure}

Let us now see what happens for a society where the agents are on the sites of a
square lattice, with periodic boundary conditions. The situation is illustrated
in Fig. \ref{fig2}. Again, two population sizes were taken, $N=1600$ and
$N=6400$, respectively. At variance with the case of the complete graph, we see
that the onset is $1/2$, as for the model of Deffuant et al. (see \cite{santo}).
This is interesting, as it reveals that the dynamics of the model of KH does not
suffice, as in Deffuant, to determine the value of the consensus threshold, but
that it is necessary to take into account the interplay between the dynamics and
the underlying graph topology. The result also confirms our expectation that 
KH becomes very similar to Deffuant when the average degree of the graph is low.

Fig. \ref{fig3} further supports our conjecture. Here the graph is a
scale free network {\'a} la Barab{\'a}si-Albert (BA) \cite{BA}, 
which has become very popular in the last years \cite{netw}.
This object can be constructed by means of a simple dynamical procedure.
One starts from a complete graph with $m$ vertices. At each iteration a new vertex 
is added and $m$ new edges are built between the new vertex and the old ones, so
that the probability of connection to some vertex $i$ is proportional to the
degree of $i$. One repeats the procedure until the network reaches the 
desired (total) number of vertices $N$. It is known that this growth process leads to a 
graph characterized by a degree distribution with a power law tail (in the limit
$N\rightarrow\infty$): the exponent
of the power law is $3$, independently of the value of the parameter $m$ (or "outdegree").
For our network we took $m=3$;  
Fig. \ref{fig3} shows the same pattern as in Fig. \ref{fig2}, so that the
consensus threshold is $1/2$ for the BA network too. 
In this way, the value $1/2$ is not relative to a special topology, but it labels
at least two classes of graphs. What do lattices and BA networks
have in common? The average degree $d$ is $4$ for the square lattice and $2m$
for the BA network, so, in both cases $d$ remains finite when the graph becomes
infinitely large ($N\rightarrow\infty$). In the case of the complete graph, instead,
$d=N-1$, so $d\rightarrow\infty$ when $N\rightarrow\infty$. Based on this fact,
we propose the following conjecture: 

\begin{itemize}
\item{There are only two possible values for the consensus threshold
    $\epsilon_c$ of the
    model of Krause-Hegselmann: if the average degree $d$ of the graph stays
    finite when the order $N$ of the graph diverges, then $\epsilon_c=1/2$;
    if $d$ diverges when $N\rightarrow\infty$, then $\epsilon_c=\epsilon_i\sim\,0.2$.}
\end{itemize}

We remark that this conjecture distinguishes between two regimes: a regime where
each agent interacts on average with a few agents (microscopic interaction), and
a regime where each agent interacts on average with a finite fraction of the
whole population (mean field interaction). It is known that the two situations are 
well separated in statistical mechanics, and that they are characterized by different
behaviours\footnote{In spin models like Ising, for instance, mean
  field theory applies for space dimensions $d\,\geq\,4$, and the relative
  critical exponents differ from those at lower dimensions.}.

\begin{figure}[hbt]
\begin{center}
\includegraphics[angle=-90,scale=0.5]{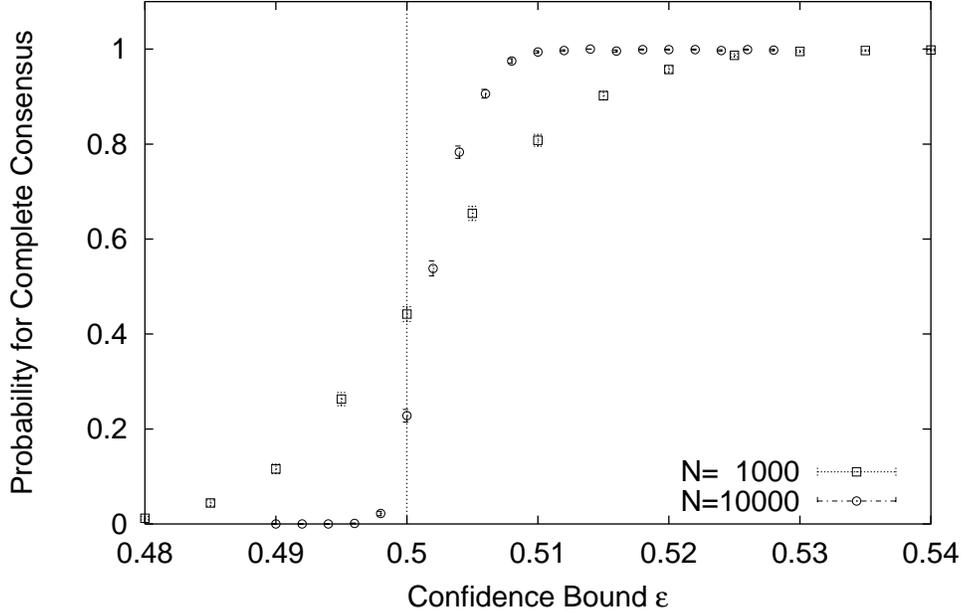}
\end{center}
\caption{\label{fig4}As Fig. \ref{fig1}, but for agents sitting on the sites of
  a random network {\'a} la Erd{\"o}s-R{\'e}nyi. Here the average degree
  $d=p(N-1)\sim pN$ is kept constant when increasing the
  population $N$ from $1000$ to $10000$ agents.}
\end{figure}

So we claim that there are two different "universality classes of graphs"
for the KH model, that we call ${\cal G}_F$ (finite degree) and ${\cal G}_I$
(infinite degree),
each of them being labeled by a special value of the consensus threshold.
The ideal way to test our conjecture would be to pass smoothly from one class of
graphs to the other,
so that we can see how the consensus threshold varies. If there were only two
possible values for $\epsilon_c$, we would expect to observe a discontinuous
variation by passing from the one to the other class. There is a special type of
graph which allows us to perform this test, the random graph of 
Erd{\"o}s and R{\'e}nyi \cite{erdos}.
It is characterized by a parameter
$p$, which is the bond probability of the vertices. One assumes that 
each of the $N$ vertices of the graph has probability
$p$ to be linked to any other vertex.
In this way, the total number of edges $m$ is 
$m=pN(N-1)/2$ and the average degree 
is $d=p(N-1)$ which 
can be well approximated by $pN$ when $N\rightarrow\infty$.
This class of graphs is especially interesting for our purposes because 
it contains both graphs in ${\cal G}_F$ and graphs 
in ${\cal G}_I$.
In fact, suppose that $p$ is fixed to some value $>0$: then 
$d=p(N-1)\rightarrow\infty$ when $N\rightarrow\infty$. On the other hand, if 
$p\rightarrow 0$ when $N\rightarrow\infty$ 
in such a way that the product $p(N-1)$ remains constant, then $d$ would
remain finite. The random graph of Erd{\"o}s and R{\'e}nyi provides us then 
a natural way to interpolate between ${\cal G}_F$ and ${\cal G}_I$.

\begin{figure}[hbt]
\begin{center}
\includegraphics[angle=-90,scale=0.5]{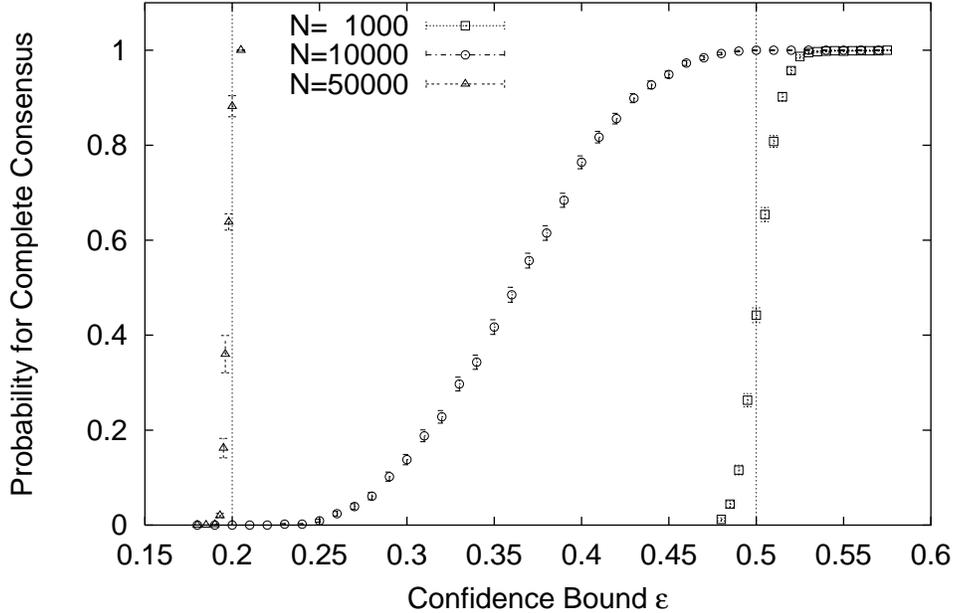}
\end{center}
\caption{\label{fig5} As Fig. \ref{fig4}, but for a fixed bond probability
  $p=0.002$. In this case the average degree $d=p(N-1)\sim pN$
  diverges when $N\rightarrow\infty$, and the consensus threshold
  jumps from $1/2$ to a smaller value which coincides, within errors, with the
  threshold obtained for the complete graph.}
\end{figure}

Fig. \ref{fig4} shows the probability for complete consensus $P_c$ as a function
of $\epsilon$ for two random graphs {\'a} la Erd{\"o}s and R{\'e}nyi, with 
$1000$ and $10000$ vertices, respectively. The bond probability $p$ is varied so
to keep the product $pN$ fixed in both cases ($pN=2$). In the limit $N\rightarrow\infty$
we would then get a graph of the class ${\cal G}_F$, and indeed we see from the
figure that the consensus threshold is $1/2$, as for the square lattice and the
BA network. The situation changes abruptly in Fig. \ref{fig5}: here we have
three random graphs, with $1000$, $10000$ and $50000$ vertices, respectively,
but now the value of the bond probability $p$ is fixed to $0.002$. For $N=1000$ we
see that the consensus threshold is close to $1/2$, but this is clearly a finite
size effect, as the system is small and has a few edges, so it is in the same
situation as the smaller graph in Fig. \ref{fig4}. However, when $N$ increases
to $10000$ we notice that the variation of $P_c$ takes place in a large  
range of $\epsilon$ values, which indicates the crossover between one regime and
the other. Finally, for $N=50000$, the variation of $P_c$ is again sharp and 
takes place in a narrow interval of $\epsilon$, which goes from $0.194$ to
$0.202$, in excellent agreement with the range we have found for the complete graph.

The last issue we would like to discuss here concerns the definition of  
consensus threshold. In all cases we have dealt with so far, the pattern of $P_c$ 
seems to approach a step function when the graph becomes infinite. In this way, 
the onset $\epsilon_c$ indicates the value of $\epsilon$ above which the system
can reach only a single stable final configuration, i.e. complete consensus.
This consideration is true as well for the model of Deffuant et al \cite{santo}.
For the opinion dynamics of KH, however, this is not the end of the story.
As a matter of fact, it can happen that consensus is
not the only possible stable state, but that it coexists with 
other stable states, even when the number of agents goes to infinity.
In such cases, how is the consensus threshold $\epsilon_c$
defined? A possible definition would be 
the value of the confidence bound above which complete consensus is a possible stable
state for the system, i. e. the value $\epsilon_c$ such that $P_c(\epsilon)>0$ 
(but not necessarily $1$) for 
$\epsilon>\epsilon_c$.

There are indeed some special graphs for which the probability for complete
consensus does not converge to a step function.
A typical example is shown in Fig. \ref{fig6}, where we plot once again 
the probability for complete consensus $P_c$ as a function of $\epsilon$.
The social topology is now a star, i. e. a graph where one vertex (the "core" of
the star, that we call $C$) is linked
to all others (the connections are the "rays" of the star). We performed
ordered updates of the agent opinions starting with the most connected central agent.
In this case, it is possible to determine analytically the expression of the 
consensus 
probability as a function of $\epsilon$ in the limit $N\rightarrow\infty$. 
The average degree of the graph is finite, as the total number of edges is
$N-1$, so $d=2(N-1)/N\rightarrow 2$ when $N\rightarrow\infty$. If we believe our
conjecture, the consensus threshold should be $\epsilon_c=1/2$.
Let us see what happens when $\epsilon>1/2$.
Suppose that the central agent $C$ gets initially the opinion $x\in [0:1]$.
There are three possibilities:

\begin{enumerate}
\item{$0<x<1-\epsilon$, all agents with opinions between $0$ and $x+\epsilon$
    are compatible with $C$, so the new opinion $s_C$ of $C$ is the average of 
all opinions in the interval $[0, x+\epsilon]$, i. e.
$s_C=(x+\epsilon)/2$;}
\item{$1-\epsilon\,\,{\leq}\,\,x<\epsilon$, all agents are compatible with $C$, so its new
    opinion is $1/2$;}
\item{$\epsilon\,\,{\leq}\,\,x<1$, all agents with opinions between $x-\epsilon$ and $1$
are compatible with $C$, so the new opinion $s_C$ of $C$ is the average of 
all opinions in the interval $[x-\epsilon, 1]$, i. e. $s_C=(1+x-\epsilon)/2$.}
\end{enumerate}

\begin{figure}[hbt]
\begin{center}
\includegraphics[angle=-90,scale=0.5]{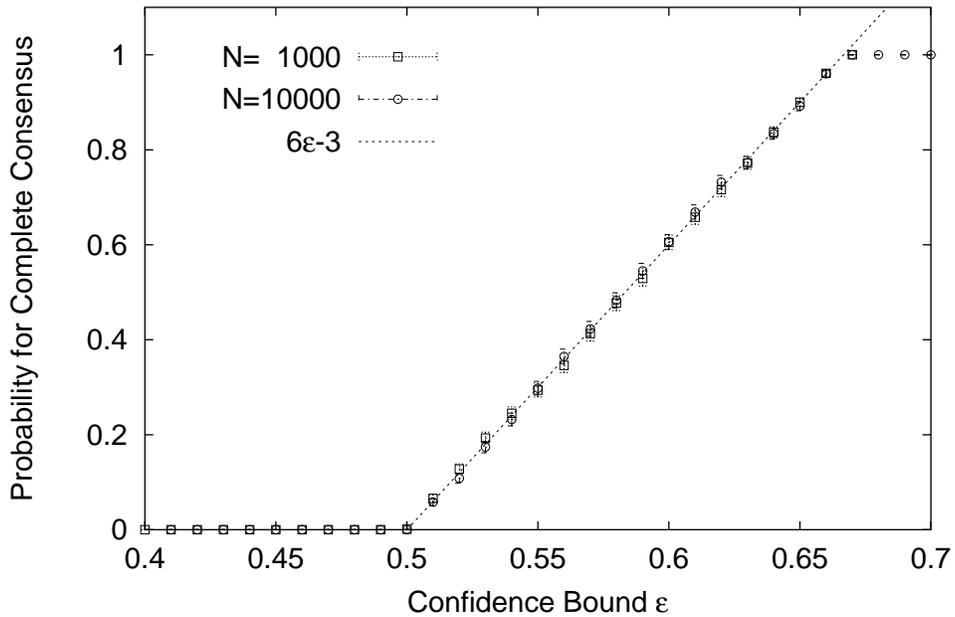}
\end{center}
\caption{\label{fig6}Fraction of samples with a single opinion cluster 
in the final configuration, for a special community 
where one agent is connected
with all the others, but the others have no further connections (star-like graph).
We performed ordered
sweeps over the whole population starting with the center of the star.
For the system to always reach 
complete consensus (i.e. with
probability $1$), $\epsilon$ must be greater than $2/3$.}
\end{figure}

The aim of this calculation is to identify those intervals of the opinion space
$[0:1]$ such that, if $x$ falls in any of them, agent $C$ will be compatible
with all other agents after the first step of the calculation. 
If this happens, in fact, $C$ will sooner or later "convince" all other people to accept 
its opinion (through successive shifts), so the system would reach complete consensus.
We indicate the size of the "consensus" intervals for $x$ in the three cases as $p_1$,
$p_2$ and $p_3$, respectively. 
In case $2$, for any $x$ in the interval $1-\epsilon\,\,{\leq}\,\,x<\epsilon$, agent
$C$ will take opinion $1/2$ after the first step and will then be compatible
with all agents of the community, as $\epsilon>1/2$. For this reason,
the size $p_2$ of the "consensus" interval for $x$ is just the length of the
whole range $[1-\epsilon,\epsilon]$, i. e. $p_2=2\epsilon-1$.
It is easy to see that, in the remaining two cases,
the sectors of the opinion space in which $x$ has to fall in order to obtain
consensus have the same size as in case 2,  
i. e. $p_1=p_3=p_2=2\epsilon-1$. 
As the opinions are uniformly distributed at the beginning of the process,  
the probability for $x$ to fall in the "consensus" intervals,
which coincides with the probability $P_c$ of having complete consensus, is
$p_1+p_2+p_3=6\epsilon-3$. This ansatz is 
represented by the skew straight line in Fig. \ref{fig6}, and it reproduces the
data very well. The data sets are actually two, corresponding to $N=1000$ and
$N=10000$ agents, respectively. Their excellent overlap shows that what 
we observe is indeed the asymptotic pattern.
We remark that $P_c>0$ for $\epsilon>1/2$, 
in agreement with our conjecture, and that $P_c=1$ 
only\footnote{As a matter of fact, the other stable configurations 
of the system for $1/2<\epsilon<2/3$ are
  characterized by one large cluster of agents with the same opinion of the
core $C$, 
and by single-agent clusters with opinions close to the extremes $0$ and/or $1$.}
for $\epsilon\,\,{\geq}\,\,2/3$.

\begin{figure}[hbt]
\begin{center}
\includegraphics[angle=-90,scale=0.5]{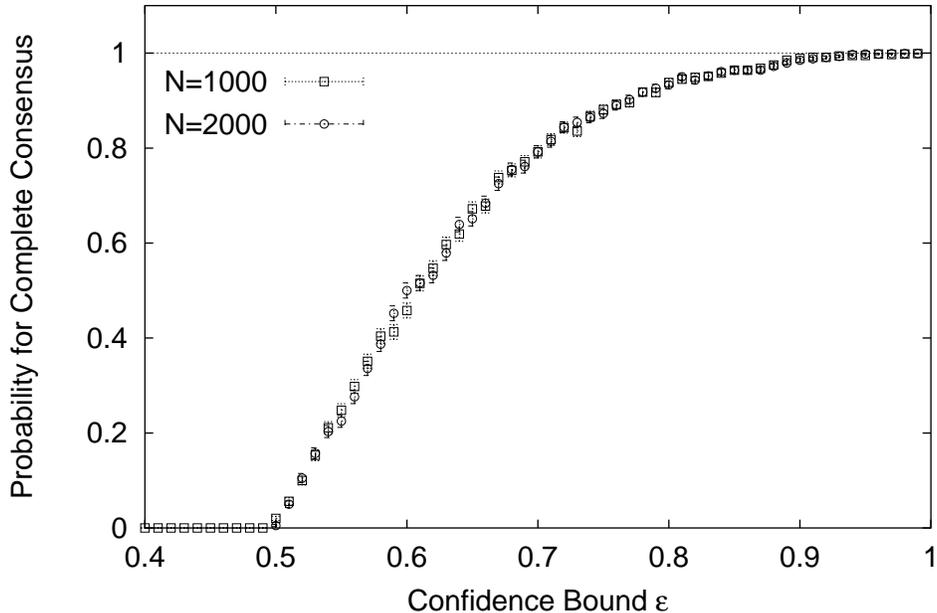}
\end{center}
\caption{\label{fig7}As Fig. \ref{fig6}, but for random updating order of the agent opinions.
In this case the pattern of the probability for complete consensus is not a
simple straight line, but a curve which reaches the value $1$ for
$\epsilon\rightarrow 1$.}
\end{figure}

For the pattern of Fig. \ref{fig6} it is essential to perform an ordered
update of the agent opinions starting with the central agent $C$. If we would
perform a random update of the agents, the situation would look quite different.
In this case, in fact, before coming to the update of $C$, the opinions of 
some finite fraction of the whole population have been varied,
and that alters the distribution of the opinions that can influence $C$, which
are now no longer uniformly distributed. This fact prevents us from repeating 
the same argument we have presented above, for which the uniformity of the
opinion distribution was crucial,
and the results are different. Fig. \ref{fig7} illustrates the new
situation. We took two population sizes, $N=1000$ and $N=2000$: their remarkable
overlap again shows that the observed pattern is the asymptotic one.
However, it is no longer a simple straight line, but a curve which attains its
limit value $1$ when $\epsilon{\rightarrow}1$. We have as well carried on simulations on
the star-like graph for the opinion dynamics of Deffuant et al.; 
in contrast with the results of Figs. \ref{fig6} and \ref{fig7}, we found that the
probability for complete consensus converges to a step function as for all other
graphs (the onset is still $1/2$).

We have found that the consensus threshold $\epsilon_c$ of the opinion dynamics of 
Krause-Hegselmann is not specific of the particular graph one uses to describe
the social relationships between individuals, but it can take only two 
possible values, $\epsilon_i\,\sim\,0.2$ and $1/2$. The criterion which distinguishes
the two possibilities is the behaviour of the average degree $d$ of the graph 
when the number of vertices $N$ goes to infinity. If $d$ stays finite,
$\epsilon_c=1/2$, as in the model of Deffuant et al.; 
if instead $d\rightarrow\infty$, then $\epsilon_c=\epsilon_i\,\sim\,0.2$.
We have tested our conjecture on different types of graphs: the complete graph, 
the square lattice, the Barab{\'a}si-Albert network, the random graphs a l{\'a} 
Erd{\"o}s and R{\'e}nyi. Further tests on star-like graphs show that 
the probability $P_c$ for complete consensus does not always converge to a step function
when $N\rightarrow\infty$, and that the updating order of the agents may influence the
final shape of the curve. We stress that for the result to hold it is necessary
that the opinion space be symmetric with respect to the center opinion $1/2$,
as in the case of the model of Deffuant et al. For modifications of the model
violating this symmetry we expect to find
different values of the consensus threshold, as 
it was recently found for Deffuant \cite{assmann}.

\bigskip

I am indebted to V. Latora for suggesting me to look for possible analogies between 
the model of KH and that of Deffuant et al. I also thank
D. Stauffer for a critical reading of the manuscript.
I gratefully acknowledge the financial support of the DFG Forschergruppe
under grant FOR 339/2-1.

\end{document}